*Kyrylo Malakhov, Aleksandr Kurgaev, Vitalii Velychko*

# MODERN RESTFUL API DLS AND FRAMEWORKS FOR RESTFUL WEB SERVICES API SCHEMA MODELING, DOCUMENTING, VISUALIZING

The given paper presents an overview of modern RESTful API description languages (belongs to interface description languages set) – OpenAPI, RAML, WADL, Slate – designed to provide a structured description of a RESTful web APIs (that is useful both to a human and for automated machine processing), with related RESTful web API modelling frameworks. We propose an example of the schema model of web API of the service for pre-trained distributional semantic models (word embedding's) processing. This service is a part of the "Personal Research Information System" services ecosystem – the "Research and Development Workstation Environment" class system for supporting research in the field of ontology engineering: the automated building of applied ontology in an arbitrary domain area as a main feature; scientific and technical creativity: the automated preparation of application documents for patenting inventions in Ukraine. It also presents a quick look at the relationship of Service-Oriented Architecture and Web services as well as REST fundamentals and RESTful web services; RESTful API creation process.

## Introduction

Databases, web sites, business applications and services need to exchange data. This is accomplished by defining standard data formats such as Extensible Markup Language (XML) or JavaScript Object Notation (JSON), as well as transfer protocols or Web services such as the Simple Object Access Protocol (SOAP) or the more popular today – Representational State Transfer (REST). Developers often have to design their own Application Programming Interfaces (APIs) to make applications work while integrating specific business logic around operating systems, or servers. This paper introduces these concepts with a focus on the RESTful APIs and presents an overview of modern RESTful API description languages (RESTful API DLs): OpenAPI Specification, RAML, and the example of modeling the schema of web API of the service for pre-trained distributional semantic models (word embeddings) processing (is a part of the "Personal Research Information System" [1] services ecosystem – the "Research and Development Workstation Environment" [2] class system for supporting research in the field of ontology engineering: the automated building of applied ontology in an arbitrary domain area as a main feature; scientific and technical creativity: the automated preparation of application documents for patenting inventions in Ukraine) with related RESTful web API modelling frameworks.

## Service-Oriented Architecture style and Web services

According to the Open Group [3] (a global consortium that develops open, vendor-neutral information technology standards), an SOA is an architectural style that supports service orientation. Service orientation is a way of thinking in terms of the outcomes of services, and how they can be developed and combined. In this definition, a service is a repeatable business activity that can be logically represented; the Open Group gives the examples: "check customer credit," and "provide weather data." Further, a service is self-contained, may be composed of other services, and consumers of the service treat the service as a black box. SOA is a distinct architectural style which is a major improvement over earlier ideas, although it includes some of the earlier ideas. Also, traditional architectural methods must be employed in order to obtain maximum benefit from using SOA.

Another definition of Service-Oriented Architecture comes from [4]: a paradigm for organizing and utilizing distributed capabilities that may be under the control of different ownership domains. It provides a uniform means to offer, discover, interact with and use capabilities to produce desired effects consistent with measurable preconditions and expectations. According to [4], the focus of SOAs is to perform a task (business function). This is different from some other paradigms, such as object-oriented architectures, where the focus is more on structure of the solution in the case of an object-oriented architecture, the focus is on how to package data inside an object. SOAs address ownership boundaries through service descriptions and service interfaces. SOA provides reuse of externally developed frameworks by providing easy interoperability between systems. Generally speaking, in order to perform a task, an SOA groups services on different systems, possibly running on different operating systems, possibly written using different programming languages. Most current SOA-based applications employ an asynchronous client/server-type architectural style – asynchronous event-driven architectural style [5]. Event-driven SOA (also known as SOA 2.0) is the current and advanced form of SOA. In this approach at present, unlike the older SOA approach where services used to be designed as predefined processes, the events generally trigger the execution of activities. The asynchronous event-driven architectural style is better for real time or proactive systems, since business processes are treated as a sequence of events, and therefore different business processes that have little relationship with each other, except for a few individual shared tasks, do not have to obey the same kind of centralized management. In an asynchronous event-driven architecture, an event message carries a state change to an event server. The event server passes these events along to the servers, possibly with value added. Servers may then generate messages for other event servers (often calls "publish/subscribe" architecture). More detailed in-depth look at the current state of SOA presented in [6, 7].

Figure 1 uses a Venn diagram to illustrate the relationship between SOA and Web services. The overlapping area in the center represents SOA using Web services for connections. The nonoverlapping area of Web services represents that Web services can be used for connections, but connections alone do not make for an SOA. The non-overlapping area of SOA indicates that an SOA can use Web services as well as connections other than Web services (the original specifications of CORBA and DCOM are examples).

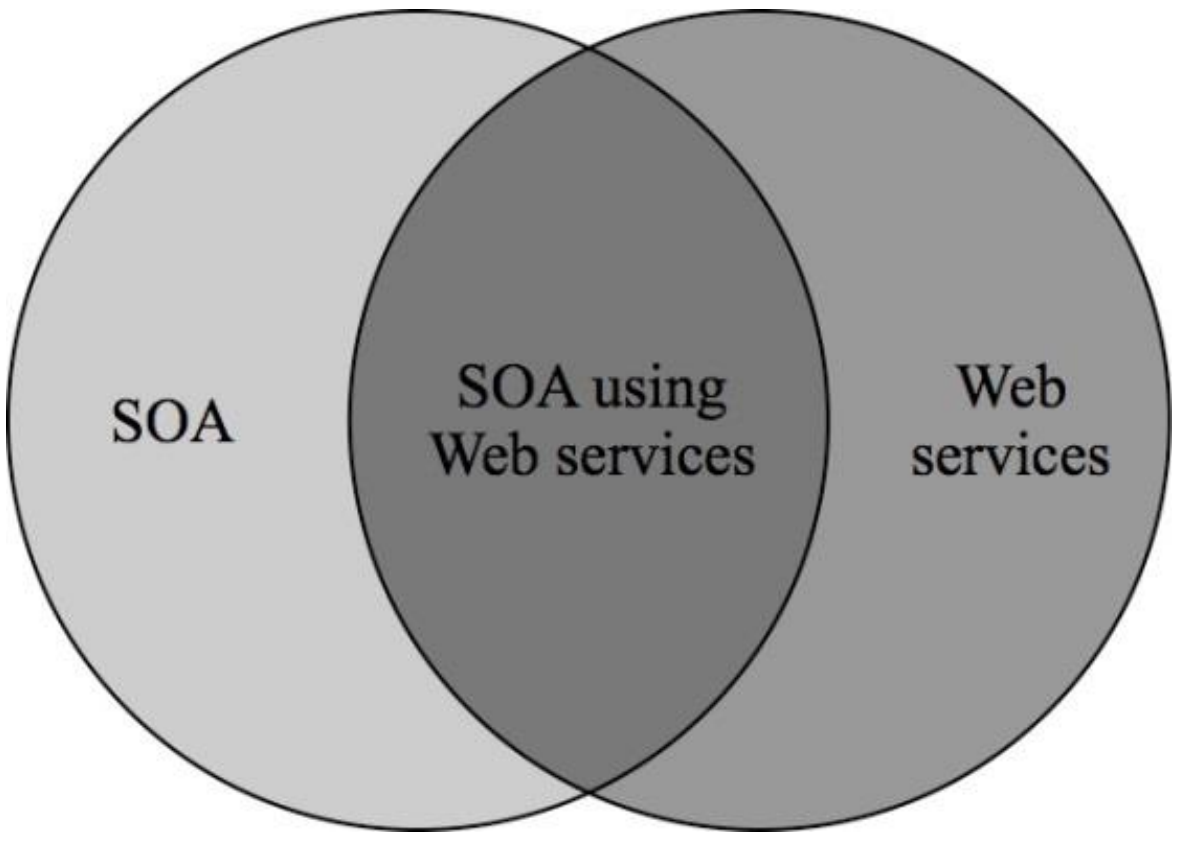

Figure 1. Relationship of Web services and SOA

Key to SOA is the identification and design of services. The idea is that services should be designed in such a way that they become components that can be assembled in multiple ways to support or automate business functions. It is not necessarily easy to properly identify and design services. When done well, the services allow an organization to quickly assemble services – or modify the assembly of services – of add or modify the support or automation of business functions. Here are basic concepts related to services [8].

- *Atomic service*. An atomic service is a well-defined, self-contained function that does not depend on the context or state of other services. Generally, an atomic service would be seen as fine grained or having a finer granularity.

- *Composite service*. A composite service is an assembly of atomic or other composite services. The ability to assemble

services is referred to as composability. Composite services are also referred to as compound services. Generally, a composite service would be seen as coarse grained or having a larger granularity.

- *Loosely coupled*. This is a design concept where the internal workings of one service are not "known" to another service. All that needs to be known is the external behavior of the service. This way, the underlying programming of a service can be modified and, as long as external behavior has not changed, anything that uses that service continues to function as expected. This is similar to the concept of information hiding that has been used in computer science for a long time.

The design challenge is to find a balance between fine-grained and coarse-grained services to minimize communication overhead yet keep the services loosely coupled.

Services are assembled to support or automate business functions. Figure 2 illustrates the assembly of services. This represents an SOA. Web services are used to connect the services in an SOA [8].

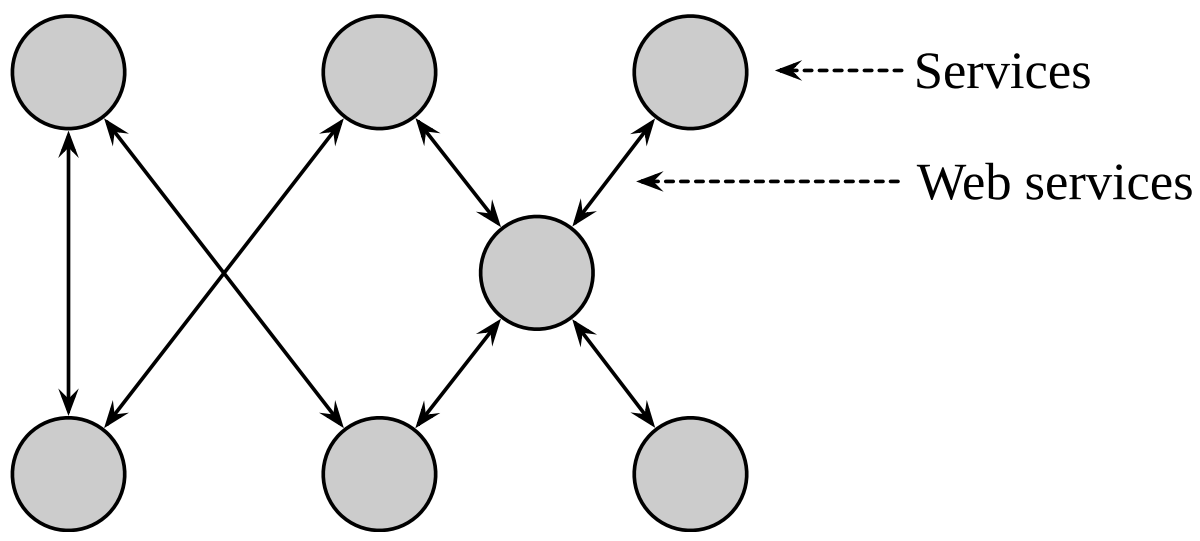

Figure 2. Assembly of services into an SOA

It is easy to imagine that we can reassemble the same services with other services to achieve a different functionality. This ability to change the assembly of services is one way that an SOA can quickly adapt to changing business needs.

## RESTful architectural style and RESTful web services

According to Fielding [9], the RESTful architectural style focuses on: "...the roles of components, the constraints upon their interaction with other components, and their interpretation of significant data elements...". He coined the term "REST" an architectural style for distributed hypermedia systems. Put simply, REST (short for Representational State Transfer) is an architectural style defined to help create and organize distributed systems. The key word from that definition should be "style," because an important aspect of REST is that it is an architectural style – not a guideline, not a standard, or anything that would imply that there are a set of hard rules to follow in order to end up having a RESTful architecture.

The RESTful architectural style consists of constraints on data, constraints on the interpretation of data, constraints on components, and constraints on connectors between components.

The RESTful architectural style possesses the following constraints [9].

*Client-Server.* The separation of concerns is the core theme of the Web's client-server constraints. The Web is a client-server-based system, in which clients and servers have distinct parts to play. They may be implemented and deployed independently, using any language or technology, so long as they conform to the Web's uniform interface.

*Stateless*. The client-server interaction is stateless. There is no stored context on the server. Any session information must be kept by the client.

*Cacheable.* Data in a response (a response to a previous request) is labeled as cacheable or non-cacheable. If it is cacheable, the client (or an intermediary) may reuse that for the same kind of request in the future. Caching response data can help to reduce client-perceived latency, increase the overall availability and reliability of an application, and control a web server's load. In a word, caching reduces the overall cost of the Web.

*Uniform Interface*. There is a uniform interface between components. In practice, there are four interface constraints: resource identification – requests identify the resources they are operating on (by a URI, for example); resource manipulation through the representation of the resource – when a client or server that has access to a resource, it has enough information based on understanding the representation of the resource to be able to modify that resource; messages are self-descriptive – the message contains enough information to allow a client or server to handle the message, this is normally done through the use of Internet Media types (MIME types); use of hypermedia to change the state of the application – for example, the server provides hyperlinks that the client uses to make state transitions.

*Layered System*. Components are organized in hierarchical layers; the components are only aware of the layer within which the interaction is occurring. Thus, a client connecting to a server is not aware of any intermediate connections.

*Code-on-Demand*. The Web makes heavy use of code-on-demand, a constraint which enables web servers to temporarily transfer executable programs, such as scripts or plug-ins, to clients. Code-on-demand tends to establish a technology coupling between web servers and their clients, since the client must be able to understand and execute the code that it downloads on-demand from the server. For this reason, code-on-demand is the only constraint of the Web's architectural style that is considered optional.

So, it's pretty clear that the RESTful web services meet the constraints of the RESTful architecture. Summarizing, a RESTful web service is client/server-based, does not store state. It accesses resources (web pages or data) located at a URL. The results of a request from client to server can be cached in the client. It has a uniform interface with self-descriptive messages, based on hypermedia. Also, the client and server aren't aware of intermediate connections between the two of them.

## RESTful API creation process – designing API and creating a schema modeling

As UI is to UX (User Experience), API is to APX (Application Programming Experience). Like optimizing for UX (User Experience) has become a primary concern in UI development, also optimizing for APX (API User Experience) should be a primary concern in API development.

The process of RESTful API creation must contain all of the following steps:

- Determining business value.
- Choosing metrics.
- Defining use cases.
- *Designing API and creating a schema model*.

A detailed description of the RESTful API creation process is presented in [8, 10, 11]. In our paper we will focus on the designing API and creating a schema model. Modeling the *schema* for your API means creating a design document that can be shared with other teams, customers, or executives. A schema model is a contract between your organization and the clients who will be using it. A schema model is essentially a contract describing what the API is, how it works, and exactly what the endpoints are going to be. Think of it as a map of the API, a user-readable and a machine-readable (automated machine processing) description of each endpoint, which can be used to discuss the API before any code is written. With a schema model, we can ensure that everyone has a shared understanding of what the API will do and how each resource will be represented when the API is complete. Each of the schema modeling languages has tools available to automate testing or code creation based on the schema model you've created. But even without this functionality, the schema model helps us have a solid understanding of the API before a single line of code is written. Figure 3 shows the API Modeling framework where you have API specifications defined and generate API documentation [12]. Also, generate server and client source code.

Next, we'll look at the specifics of two of the main schema modeling frameworks and markup languages:

- RESTful API Modeling Language (RAML), which supports Markdown.
- OpenAPI specification (OpenAPI) format (previously Swagger), which supports JSON and YAML.

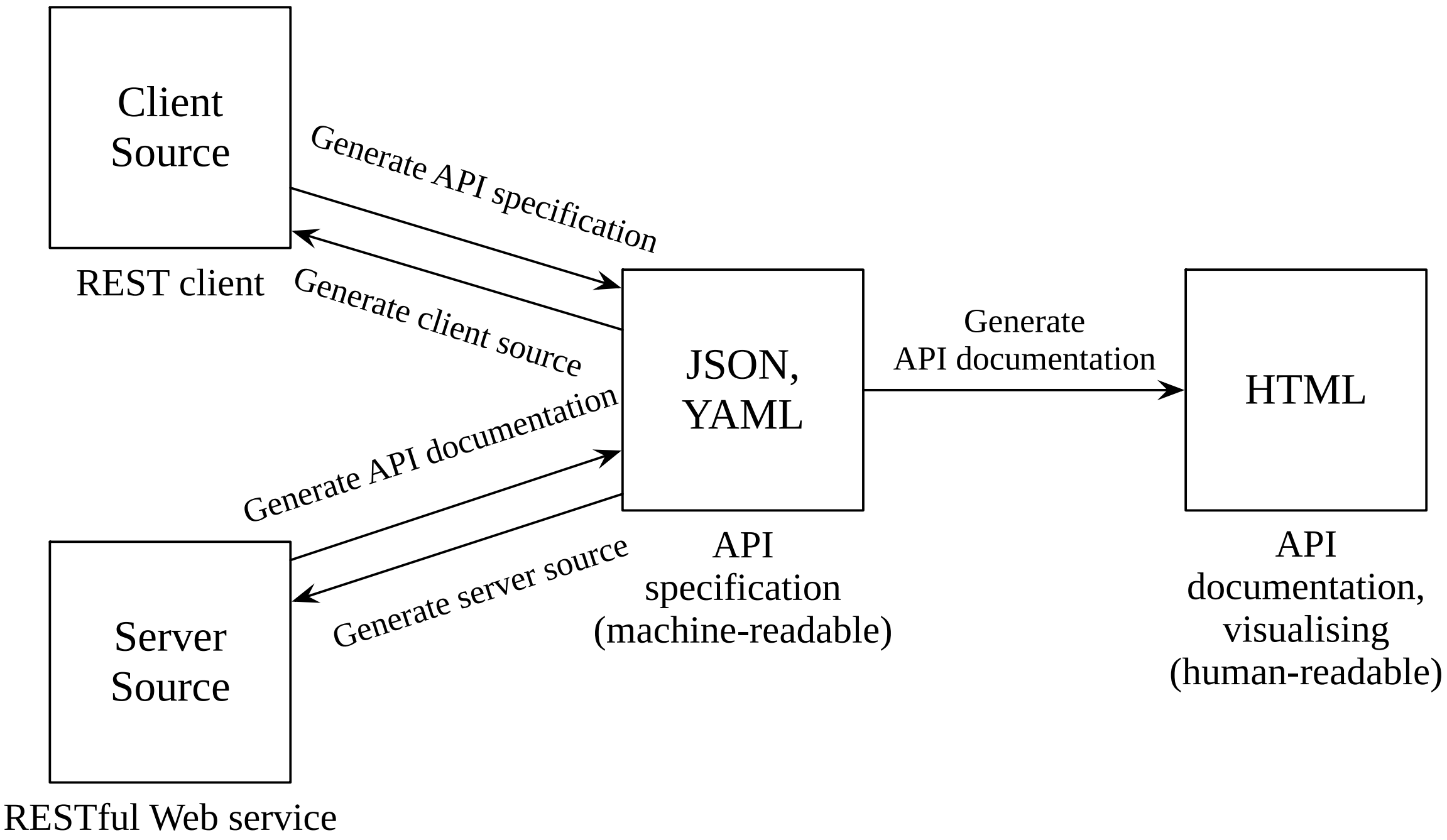

Figure 3. API modelling

## RAML and OpenAPI: an overview

The RESTful API Modeling Language (RAML) [13] is a concise, expressive language for describing RESTful APIs. Built on broadly used standards such as YAML (YAML stands for Yet Another Markup Language, and is a generic specification language) and JSON, RAML is a non-proprietary, vendor-neutral open spec. RAML was created around the notion of design-first development [12]. Although all of the specification languages can be used this way, RAML was designed this way from the outset. It makes it easy to create a code development life cycle that supports the development of APIs that meet your business goals and use cases. The RAML website [14] has good documentation, including strategies, best practices, and practical instruction. You'll find a basic tutorial for the RAML language itself at [14]. RAML has good online modeling tools, also, it has been open-sourced along with tools and parsers for common languages. The development of RAML will be overseen by a steering committee of API and UX practitioners, and there is an emerging ecosystem of third-party tools being developed around RAML [15]. Consider the pros and cons of RAML [16]. *Pros*: single specification to maintain; strong, visual-based integrated development environment and online tooling with collaboration focus; allows for design patterns; easy to get started. *Cons*: lacks strong documentation and tutorials outside of specification; limited code reuse/extensions; multiple specifications required for several tools, including dev and QA; poor tooling support for newer versions.

The best way to get started with RAML is to use the *RAML API Designer* with free account on the Anypoint system, where MuleSoft maintains its RAML specific tools [17]. RAML excels at supporting the entire API's lifecycle. It provides a balance between developer tooling and technical writers without taking away from one or the other. It also is the fastest framework to ramp up your project. MuleSoft maintains some open source tools that can extend and improve experience with a RAML specification. The API Designer that helps you design your schema from the ground up. An API Console graphical user interface is available that displays the structure and patterns and creates interactive documentation. The API Notebook provides a way to use JavaScript to test and explore APIs and create Markdown versions of the API to share on GitHub. You'll find hundreds of additional

RAML tools at GitHub and on the [13] website, which can help you create and leverage the schemas you build.

The *OpenAPI Specification* OpenAPI, originally known as the Swagger Specification, is a specification for machine-readable interface files for describing, producing, consuming, and visualizing RESTful web services. Originally part of the Swagger framework [18], it became a separate project in 2016, overseen by the OpenAPI Initiative, an open source collaborative project of the Linux Foundation [19]. Swagger and some other tools can generate code, documentation and test cases given an interface file. OpenAPI was one of the earliest schema modeling frameworks available, and it has gone through a few revisions. Version 3.0 is the most recent one as of this writing. During the development of the various versions, they've incorporated many of the best practices uncovered by the other two languages, and OpenAPI remains one of the innovative frameworks available. OpenAPI supports both JSON and YAML for its schema markup. Consider the pros and cons of OpenAPI [16]. *Pros*: a large community and support-base; high adoption rate, meaning lots of documentation; strong framework support; has the largest language support of any opensource framework. *Cons*: requires multiple specifications for some tools, including dev and QA; doesn't allow for code reuse, includes, or extensions; lacks strong developer tools; requires schemas for all responses.

OpenAPI has a very strong modeling language for defining exactly what's expected of the system – very useful for testing and creating coding stubs for a set of APIs.

In comparison to one another, both OpenAPI and RAML are very capable, compatible with many languages.

- Both offer compatibility in: .NET, Go, Haskell, Java, JavaScript, Node.js, PHP, Python, Ruby, Scala.
- OpenAPI's additional capabilities: Clojure, Coldfusion, D, Eiffel, Erlang, Groovy, and Typescript.
- RAML's additional capabilities: Elixer and Pearl.

Both languages are strong and able to produce excellent APIs despite their differences. Their key differences are what can help you determine which is best for your business.

OpenAPI's best features are its strong documentation and compatibility with lesser used languages. It provides a fast setup and a large support community. The big takeaway for OpenAPI is that it is designed as a bottom-up specification. OpenAPI specifies the behavior which affects the API to create more complex, interlocking systems.

RAML excels at supporting the entire API's lifecycle. It provides a balance between developer tooling and technical writers without taking away from one or the other. It also is the fastest framework to ramp up your project. The main difference between the two is that RAML is a top-down specification, meaning it breaks down the system and explains the behavior of the various subcomponents.

The main characteristics of both RESTful API DLs are presented in the comparative table.

There are, of course, alternatives. Two of the most popular are WADL [20] and Slate [21]. Each have their own caveats, of course. WADL is incredibly time consuming to create descriptions with, and the linking methodology leaves much to be desired when compared to any of the three specifications discussed throughout this article. Slate, similarly, has the caveat of having untested or unproven approaches due to the relatively small userbase, despite the fact that it handles documentation much like API Blueprint [22] does, and generates a pretty interface for it all.

These alternatives are interesting, to be sure, but their low adoption rates, issues inherent to their structure, and fundamental caveats make a potentially unstable bet. With many strategies in the modern IT workforce focusing heavily on rapid development and deployment, untested approaches have the distinct possibility of massively lowered quality as the demand rises exponentially.

As part of the development of the "Personal Research Information System" [1, 2], the API schemas of its services was modeled with OpenAPI, in particular, the schema model of web API of the service for pre-trained distributional semantic models (word embeddings) (DSM) processing. With this web service API is possible to: calculate semantic similarity between pair of terms (including multiple-word terms, one-word terms, words) within the chosen DSM; compute a list of nearest semantic associates for terms (including multiple-word terms, one-word terms, words) within the chosen DSM; find the center of lexical cluster for a set of terms (including multiple-word terms, one-word terms, words) within the chosen DSM; calculate semantic similarity between two sets of terms (including multiple-word terms, one-word terms, words) within the chosen DSM.

Table. Comparison of modern RESTful API DLs and frameworks

| Description Language | RAML | OpenAPI | WADL | Slate |
|---|---|---|---|---|
| Software license | Apache 2.0 | Apache 2.0 | CDDL 1.1 | Apache 2.0 |
| Format | YAML (Markdown) | YAML, JSON | XML | Markdown |
| Open source | yes | yes | yes | yes |
| Commercial offering | yes | yes | no | no |
| Sponsored by | Mulesoft, Cisco, VMware, Paypal, AngularJS, Box | Open API Initiative, Google, IBM, Mcrosoft | Oracle | - |
| Current release | 1.0 | 3.0 | - | 2.3.1 |
| Design strategy | API-first | Existing API | Existing API | Existing API |
| References | http://raml.org | http://swagger.io | https://github.com/javaee/wadl | https://github.com/lord/slate |
| Code generation | yes | yes | no | no |
| Documentation | yes | yes | yes | yes |
| Visual-based IDE | yes | yes | no | yes |
| Online IDE | yes | yes | no | no |
| Editors | API Workbench (IDE based on Atom) | Swagger Tools (editor, codegen, UI) | no | Local web editor |

The source code and the service API schema model description are available via GitHub repository [23].

## Conclusion

OpenAPI as well as RAML have very much in common. Projects relying on the extensive language support and tool integrations will tend to OpenAPI. But if the language support is not crucial as implementations are foremost done in standard languages such as Java, RAML is an equivalent option. OpenAPI and RAML both have a large community and are backed by market leaders, so it will never be wrong choosing one of them for API documentation.

Recently, several APIs contributors (members of 3Scale, Apigee, Capital One, Google, IBM, Intuit, Microsoft, PayPal, Restlet and SmartBear) have announced the Open API Initiative [19], which aims at standardizing the way REST APIs are described. This initiative will extend the Swagger specification and format to create an open technical community where members can easily contribute to building a vendor-neutral, portable and open specification for providing metadata for RESTful APIs. We hope this initiative will also promote and facilitate the adoption and use of a standard API Description Language.

## Література

***About the authors:***

*Kyrylo Malakhov*,
Junior Research Fellow.
38 Ukrainian publications,
3 International publications,
H-index: Google Scholar – 4.
http://orcid.org/0000-0003-3223-9844.

*Aleksandr Kurgaev*,
Doctor of Technical Science,
Professor, Leading Researcher of Department 205 at Glushkov Institute of Cybernetics.
Author of more than 240 scientific works, including 8 monographs,
100 Patents and Author's Certificates for innovations and useful models.
H-index: Google Scholar – 5, Scopus – 2.
http://orcid.org/0000-0001-5348-2734.

*Vitalii Velychko*,
PhD, assistant professor, Senior researcher.
73 Ukrainian publications,
25 International publications,
H-index: Google Scholar – 7, Scopus – 1.
http://orcid.org/0000-0002-7155-9202.

***Affilation:***

V.M. Glushkov Institute of cybernetics of National Academy of Sciences of Ukraine,
40 Glushkov ave., Kyiv,
Ukraine, 03187.
Phone: (+38) (044) 526 3348.
Email: aduisukr@gmail.com